\documentclass[11pt, a4paper]{article}
\usepackage{jheppub}
\usepackage{graphicx}
\usepackage{epsfig,amsmath,amsthm}

\newcommand{\be}{\begin{equation}}
\newcommand{\ee}{\end{equation}}
\newcommand{\bea}{\begin{eqnarray}}
\newcommand{\eea}{\end{eqnarray}}
\def\bse{\begin{subequations}}
\def\ese{\end{subequations}}
\newcommand{\IR}{\mathbb{R}}
\newcommand{\IC}{\mathbb{C}}

\def\IZ{\relax\ifmmode\hbox{Z\kern-.4em Z}\else{Z\kern-.4em Z}\fi}

\newcommand{\non}{\nonumber \\}

\def\half{\frac{1}{2}} 

\def\del{{\partial}}

  \def\hd{{\hat d}}
 
\def\al{\alpha} 
  \def\eps{\epsilon}

\def\presub{\vspace{.5cm} \noindent}

\def\bi{\begin{itemize}} \def\ei{\end{itemize}}

\def\Schw{Schwarzschild }
\def\({\left(} \def\){\right)}
\def\[{\left[} \def\]{\right]}

\title{\center{Classical
 Effective Field Theory \\  for \\ Weak Ultra Relativistic Scattering}}

\author{Barak Kol\\
 {\it Racah Institute of Physics, Hebrew University\\
 Jerusalem 91904, Israel}\\
 {\tt\href{mailto:barak_kol@phys.huji.ac.il}{barak\_kol@phys.huji.ac.il}}}

\abstract{Inspired by the problem of Planckian scattering we describe a
classical effective field theory for weak ultra relativistic
scattering in which field propagation is
instantaneous and transverse and the particles' equations of
motion localize to the instant of passing. An analogy with the
non-relativistic (post-Newtonian) approximation is stressed. The
small parameter is identified and power counting rules are
established. The theory is applied to reproduce the leading
scattering angle for either a scalar interaction field or
electro-magnetic or gravitational;
 to compute some subleading corrections,
 including the interaction duration; and to allow for
 non-zero masses. For the gravitational case we present an
appropriate decomposition of the gravitational field onto the transverse plane together with
its whole non-linear action. \\
 On the way we touch upon the relation with the eikonal
approximation, some evidence for censorship of quantum gravity,
and an algebraic ring structure on 2d Minkowski spacetime.}

\begin{document}

\maketitle

\section{Introduction}

The effective field theory (EFT) approach to General Relativity
(GR) \cite{GoldbergerRothstein1} borrows ideas from effective
quantum field theories and applies them to the two-body
post-Newtonian
 dynamics in classical General Relativity.\footnote{See
\cite{DamourFarese} for early precursors of the EFT approach to
GR.} In addition to contributing insight and a fresh perspective
the approach allowed to push forward the state of the art
\cite{PortoRothstein}
 by computing for the first time the next to
leading spin1-spin2 interaction in the effective two-body
action.\footnote{Even if imperfectly, since it was missing certain
contributions found in \cite{SteinhoffHergtSchaefer} using
Hamiltonian methods and also found later in \cite{Porto:2008tb} to
arise from indirect contributions in the EFT method. See
\cite{Levi:2008nh} for a derivation using Non-Relativistic Gravitational (NRG) fields.}

Most of the ensuing work on the EFT approach centered on the
post-Newtonian approximation -- see for example \cite{NRGR}.
However, the EFT approach has a very wide applicability range,
namely whenever a field theory (not necessarily of gravity)
contains two widely separated length (or time) scales
\cite{CLEFT-caged}. Thus far some such applications appeared: a
post-Newtonian theory for caged black holes (black holes localized
within a compact dimension) \cite{caged}; black hole internal
degrees of freedom at low frequencies \cite{dissip}; new
observations on finite size corrections to the radiation reaction
force in classical electrodynamics \cite{Galley:2010es}; some
results on the gravitational two body problem in the extreme mass
ratio limit, see \cite{GalleyHu1} and continuations of it; and
finally see \cite{Cannella:2009he} for some tests of GR.

In this work we shall present a new classical effective field
theory, namely for weak ultra-relativistic scattering, in the
context of Einstein's gravity or more generally any other field
theory.

We start in subsection \ref{planck-subsection} by reviewing past
work on Planckian scattering including \cite{'tHooft87,ACV} as a
concrete challenge for quantum gravity to address. This part
serves as motivation for our work but the ensuing parts are independent
of it.

\begin{figure}[t]
\begin{center}
\includegraphics[width=7cm]{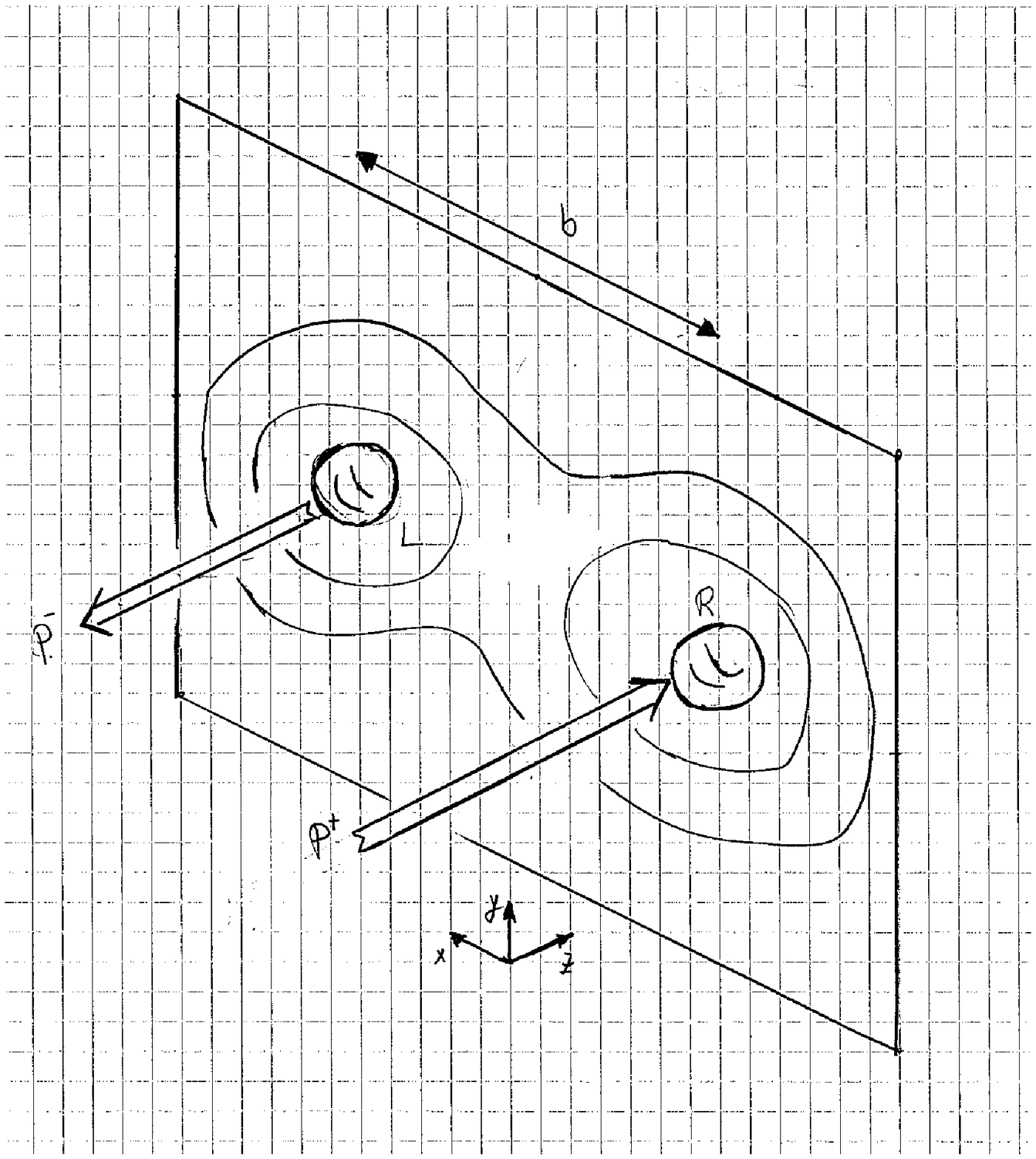}
\includegraphics[width=8cm]{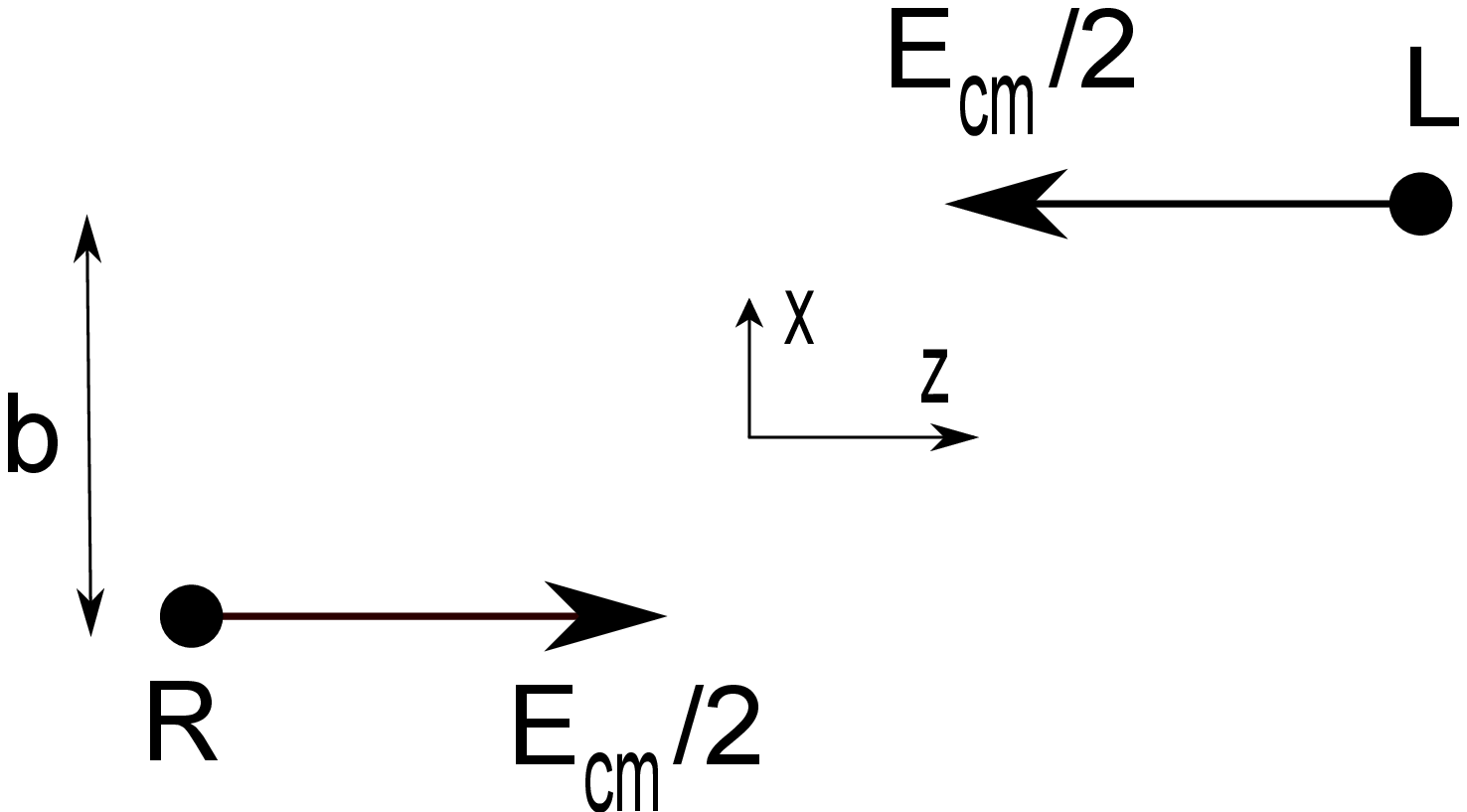}
\caption{The geometry of the scattering process. Left: 3d geometry
at the moment of passing, including equipotential lines. Note that
the field lines are Lorentz compressed to the transverse plane.
Right: top view. The initial condition parameters are the center
of mass energy $E_{cm}$ and the impact parameter $b$. The right
moving particle, namely the one moving in the positive $z$
direction, is denoted by R, while the other is denoted by L.}
\label{Geometry}
\end{center}
\end{figure}

In section \ref{set-up-EFT-section} we set up the effective field
theory and determine its central Feynman rules. The main ingredient
is to set the field propagator to be \be
 \frac{1}{k_\perp^2} \label{gen-prop} \ee
This means that the field is instantaneous and transverse. The
intuition is that while a particle at rest has a spherical
configuration of field strength lines emanating from it, an ultra
relativistic boost Lorentz contracts the longitudinal direction,
thereby confining the field lines to a transverse pancake attached
to the particle's ``nose'' (see figure \ref{Geometry}).
Accordingly the particles interact only at the moment of passing
when they each intersect with the other pancake, and the equations
of motion degenerate to that instant.

We conclude this section by computing the leading momentum
transfer during scattering for several cases of interaction
fields: scalar, electromagnetic and gravitational, reproducing the
literature value in the latter case.\footnote{The other cases may also exist in the literature.} On the way we make use of the
light-cone formalism where the dynamics of each particle is nearly
Galilean, we define effective ultra-relativistic charges and
we define a conserved global charge for the interaction fields. We observe
that this theory is analogous to the non-relativistic EFT
\footnote{In the gravitational case the non-relativistic
approximation is known as post-Newtonian, and the associated EFT
\cite{GoldbergerRothstein1} is known as Non-Relativistic GR
(NRGR).}
 only in one fewer dimension: the fields propagate
instantaneously and the particle dynamics is nearly
non-relativistic.

In section \ref{eff-action-section} we proceed to compute some
higher order corrections, present even for linear fields, which
are due to non-linearities in the \emph{solutions} of the
equations of motion, where in particular retardation effects must
be considered. Setting-up a perturbative theory for
ultra-relativistic particles and using the two-body effective
action (where the field is integrated out) allows us to find the
leading order longitudinal momentum exchange. It comes at second
order in the small parameter, while the transverse momentum
transfer is of first order, thereby confirming the assumed
hierarchy of scales which defines the EFT.

In section \ref{interaction-time-section} we choose a different time for initial conditions thereby changing our ``regularization condition'' and hence ``resumming'' and simplifying the perturbation series. This ingredient enables us to proceed to the third order and in particular to resolve the instantaneous interaction and to obtain its duration. This result demonstrates the utility of our formalism.

In section \ref{mass-corr-section} we compute corrections due to
non-zero masses, which are controlled by another small parameter.
Finally in section \ref{discussion-section} we summarize our
results and discuss open questions. The appendices contain some
conventions and calculations, as well as an algebraic curiosity.

In section \ref{URG-section} we
describe the dimensional reduction
of the gravitational field  onto the transverse
plane and its whole non-linear action. The analogue reduction for electromagnetic case is
considerably simpler and is treated in section
\ref{eff-action-section}.

The EFT in this paper is related to the eikonal approximation in
quantum field theory. Actually the latter is the semi-classical
approximation corresponding to this EFT. Some of our results
including an effective field theory for Planckian scattering
appeared elsewhere including in Amati-Ciafaloni-Veneziano[93]
\cite{ACV93}; however, we have a different approach: our EFT is
purely classical, we define the small parameter and establish
power counting rules, we analyze fields other than gravity, and we
compute various corrections including the interaction time.

We note that the ultra-relativistic limit is part of the
post-Minkowski approximation (see \cite{postMink} and references
therein) where one considers perturbations around a flat
space-time due to bodies with arbitrary velocities, possibly
relativistic but not ultra-relativistic. However, for our more
restricted velocity range we are able to construct an EFT which
allows to compute arbitrary higher order corrections, while in the
post-Minkowski case so far only a first order correction was
demonstrated to be possible.

\subsection{Planckian scattering and classical dominance}
\label{planck-subsection}

In order to explore quantum gravity it is natural to consider a
gedanken scattering experiment where two gravitationally
interacting bodies are scattered (see figure \ref{Geometry}),
especially when the center of mass energy is Planckian or higher
\cite{Dray-`tHooft,'tHooft87,ACV,ACV93,ACV2008}. For simplicity we
may assume the particles to be light-like (massless) and
point-like. Later we shall study corrections when the masses are
non-zero but still much smaller than the energies. We shall also
discuss finite-size corrections.

The initial conditions are specified by two parameters \be
E_{cm}, ~b \ee
 the center of mass energy, and the impact parameter.

\begin{figure}[t]
\begin{center}
\includegraphics[width=12cm]{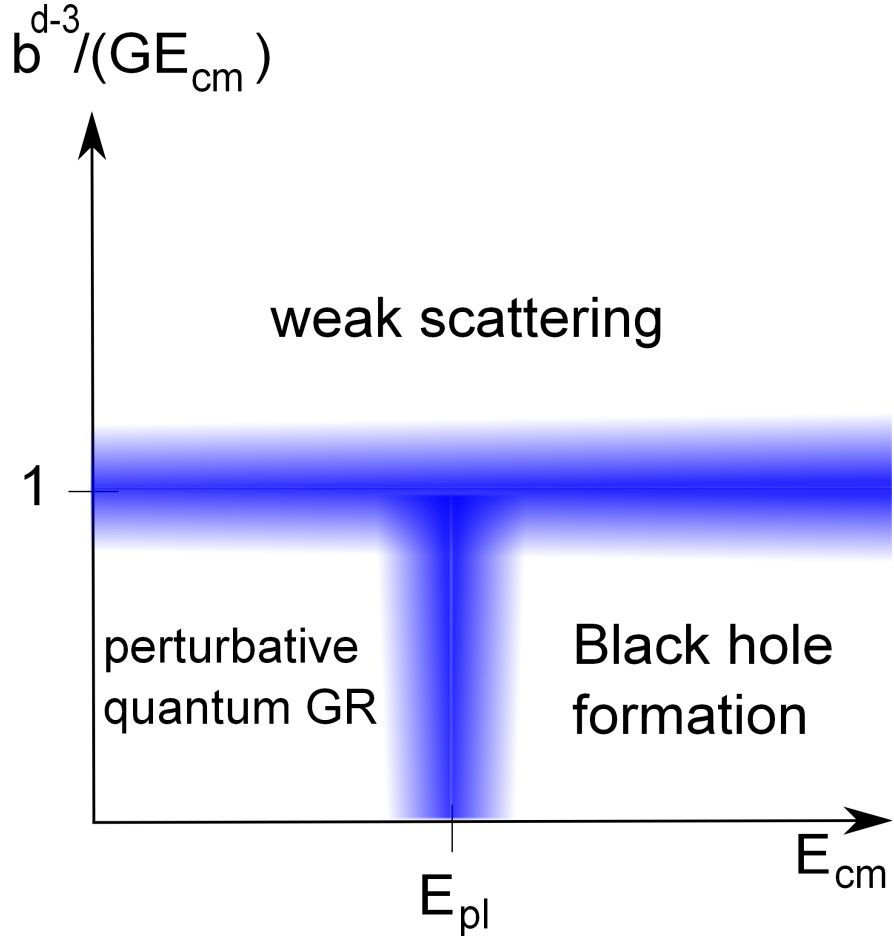}
\caption{The parameter space for trans-Planckian scattering,
demarcating the boundaries between three theoretical regions.}
\label{parameter-space}
\end{center}
\end{figure}

Let us discuss the physics as a function of this parameter space
(see figure \ref{parameter-space}). For generality we consider a
general space-time dimension $d$.
For $b=0$ we may compare $E_{cm}$ with the Planck energy $E_{Pl}
:\simeq \hbar/\(\hbar G\)^{1/(d-3)}$ (where $G$ is Newton's
gravitational constant). For large $E_{cm}$ we expect a
semi-classical black hole to form, followed by a long period of
Hawking evaporation, while for small $E_{cm}$ tree-level quantum
GR would be valid (loops are suppressed by $G E_{cm}^{d-2} \sim
\(E_{cm}/E_{Pl}\)^{d-2}$).\footnote{Actually in the quantum regime
the impact parameter is not a good parameter due to uncertainty,
and it should be replaced say by the Mandelstam variable $t$.}

For large $b$ on the other hand, we should compare $b$ with the
would-be \Schw radius $R(E_{cm}) := \(G E_{cm}\)^{1/(d-3)}$ which
is the only other parameter with length dimension. Accordingly
large $b$ means $b \gg R(E_{cm})$. In this regime the interaction
is dominated by a single field exchange, so having no loops this
is essentially classical scattering.

It is interesting to consider varying $b$ from large to small,
while keeping $E_{cm}$ fixed and large in Planck units. For large
$b$ we have a small attractive deflection. As $b$ is reduced the
deflection angle increases together with the deflection time.
While (classical) gravitational radiation  is weak for weak
deflection, it too increases as $b$ is lowered. At some critical
(or marginal) $b=b_c$ a black hole forms. By dimensional analysis
$b_c$ must be order one in units of $R(E_{cm})$. We see no reason
why the marginal black hole mass would not be zero. That means
that at $b_c$ all the  incoming energy is emitted in the form of
gravitational radiation. From general considerations the
scattering angle and time diverge in this limit\footnote{The
generic dependence of these quantities in such a marginal event
would be  $\propto \log (b-b_c)$. The simplest example for that is
a non-relativistic, non-quantum particle scattering in 1d off a
potential maximum such that the particle's energy is close to that
of the peak.}. Recently numerical GR simulations succeeded for the
first time in simulating a black hole creation in this process
\cite{ChoptuikPretorius} (and references therein). Incidentally we
note that the simulations indicate that the marginal small black
hole is very close to extremal spin.

For $b<b_c$  the black hole mass is non-zero, and the dependence
on $b_c-b$ defines a certain characteristic exponent (and the same
holds for its angular momentum). After the appearance of a black
hole horizon, the black hole relaxes to its stationary form
through the classical process of ``ringdown'' emission of
gravitational waves. The ringdown radiation is a smooth
continuation of the inspiral radiation.

As $b \to 0$  we know that the colliding light-like shock fronts
do not interact until they reach $z=0$. Being at sub-Planckian
distance this collision is outside the validity of classical GR,
yet we expect most of the energy to be converted to a black hole,
and therefore the results of this quantum gravity process would
remain hidden behind a horizon. At infinity one observes the
Hawking radiation which is semi-classical apart for the last burst
of one Planck mass which is an essential quantum gravity effect.
Due to Hawking evaporation black hole formation does not
necessitate a genuine phase transition (since for all $b$
eventually all energy returns to infinity) but rather it is
probably a ``cross-over''.

Altogether we observe that full quantum gravity is not among the
theories tiling the parameter space of figure
\ref{parameter-space}, being needed perhaps only at the boundary
at $E \sim E_{pl}$ and small $b$. Hence it appears that the
horizon conspires to hide such effects in a way that can be called
\emph{censorship of quantum gravity}.

In particular, the high energy limit is dominated by classical
gravity. This phenomenon of ``classical dominance'' was stated by
't Hooft in \cite{'tHooft87}. There the line of argument is
somewhat different. The author analyzed the collision of two
point-like and light-like gravitational shock waves. These shock
waves are described by the Aichelburg-Sexl metric
\cite{Aichelburg-Sexl} which was shown to be equivalent to a
non-trivial gluing of a pair of flat half space-times -- the one
before the shock and the one after it \cite{Dray-`tHooft}.
Considering the evolving wave function of a particle as it passes
through the shock wave carried by another, the author computes the
scattering amplitude, and finally concludes that it is dominated
by classical physics.

Other interesting aspects of Planckian scattering are discussed in
\cite{'tHooft90,Verlinde91}.

While the whole high energy range is covered by classical GR,
there is a big difference between large and small $b$: for large
$b$ we have weak scattering and a perturbative descriptions is
possible, while for small $b$ a black hole is created in what is
surely a non-perturbative process. Accordingly we may hope to find
a novel effective field theory for large $b$, but not for the more
interesting non-perturbative regime, which yields only to a full
numerical simulation. Yet, a perturbative analytic theory does
serve several purposes: it can test and validate the numerics, and
it provides insight into the problem, especially into the
dependence on parameters. It could even be used to provide
indications for the non-perturbative regime via extrapolation.
Finally and most importantly we wish to set-up a novel classical
effective field theory. Hence from now on we shall study only the
weak scattering limit.


\section{Setting-up the effective field theory}
\label{set-up-EFT-section}

Consider the two-body interaction of two ultra-relativistic
bodies, see figure \ref{Geometry}, as described in the previous
subsection. We are interested in the perturbative limit of weak
scattering, namely large impact parameter $b$. Within this limit
gravitation is not special and a scalar or vector field will also
present similar EFT's. Actually the scalar case suffices to derive
the special form of the EFT propagator, and so we begin with it.

The total action for our two particles (denoted R,L for right and
left moving) coupled to a scalar interaction field
$\phi=\phi(x^\mu)$ is \be
 S_{tot} = S[\phi] + \sum_{A=R,L} S_A[X_A,\hat{e}_A; \phi] \label{total-scalar-action} \ee
where the field action is \footnote{Recall that with this
normalization of the kinetic term the propagator in the 3d
Euclidean configuration is $+1/r$.} \be
 S[\phi] =  \frac{1}{8\pi} \int d^d x ~\del_\mu \phi ~\del^\mu \phi  ~. \ee
 The action for each particle is \be
 S = S_{kin}[X,\hat{e}] + S_{int}[X,\hat{e};\phi] \label{particle-action1} \ee
where the particle kinetic action and the interaction action are
\bea
 S_{kin}[X,\hat{e}] &=& -\frac{m}{2} \int ds ~\dot{X}^2/\hat{e} -\frac{m}{2} \int \hat{e}\,
 ds \non
 S_{int}[X,\hat{e};\phi] &=& - q \int \phi \, \hat{e}  ds \nonumber 
 \eea
where $m,\,q$ are the mass and scalar charge of each body,
$X^\mu=x^\mu(s)$ is the particles trajectory as a function of its
world-line parameterization $s$ and \be
 \dot{X}^2 := \eta_{\mu\nu} \dot{X}^\mu \dot{X}^\nu \equiv 2 \dot{X}^+ \dot{X}^- - \dot{\vec{X}}^2 ~,\ee
 where the dot denotes a derivative with respect to $s$, and in this paper arrowed vectors always denote
transverse space vectors \be
 \vec{X} \equiv X_\perp \equiv X^i ~ i=1,\dots,d-2
 \ee
 -- see appendix \ref{app-conventions} for a summary of our light
 cone conventions.
Finally $\hat{e}=\hat{e}(s)$ is an auxiliary field, the world-line
metric, which guarantees reparameterization invariance by the
transformation law $\hat{e} \to \hat{e}'=\hat{e}\, ds/ds'$ under
$s \to s'(s)$. Later we shall redefine the variable $\hat{e}$ for
convenience.

\presub {\bf The field propagator - the main observation}. Recall
that when constructing an effective field theory for a
non-relativistic approximation such as the Post-Newtonian
approximation of gravity \cite{GoldbergerRothstein1} one considers
first field modes whose typical length scale is the inter-body
separation. These modes are carried along with the particles and
are called potential modes. Since by definition the velocities are
low in this limit the potential modes satisfy $\del_t \ll \del_x$,
namely their temporal gradients are negligible with respect to
spatial gradients. Accordingly one decomposes the quadratic bulk
action as follows $4 \pi\, S = \half \int dx \del_\mu \phi
~\del^\mu \phi = -\half \int dx \, \left| \del_i \phi \right|^2 +
\half \int dx\, \left( \del_t \phi \right)^2, ~ i=1,\dots,d-1$
 -- the space-gradients constitute the unperturbed
action and define the propagator, while the time-gradients are a
perturbation and define a vertex. Apart from the potential modes,
there are radiation modes which are generated at higher orders.
The radiation modes have the same typical frequency as the
potential modes, but being on-shell they have a much longer
typical wavelength.

Similarly in the current ultra-relativistic case we shall find
that the interaction field decomposes into potential and radiation
modes, and that for potential modes certain components of $k_\mu$,
the wavenumber vector, dominate.

Consider computing the effective action for two bodies, in the
straight line approximation, namely the one-field exchange
diagram. It can be computed by evaluating (minus) the bulk kinetic
term for the field.
Substituting \be
 \phi=\phi_R + \phi_L \ee
 into the action (\ref{total-scalar-action}) where $\phi_A$ is the field sourced by particle $A=R,L$, we have \be
 4 \pi\, S =  \int dx ~ \del_\mu \phi_R \, \del^\mu \phi_L ~. \label{kinetic-action} \ee

 When we compute (\ref{kinetic-action}) in $k$ space we essentially evaluate the integral \be
 \int dk\, k^2 ~\phi_R(k) ~\phi_L(-k) \ee
Since $\phi_R$ is independent of $z^+$ then $\phi_R(k) \propto
\delta(k^-)$ and vice versa for body L (since $x \cdot k \supset
z^+ k^- + z^- k^+$).
 The product integrand is now localized in
\emph{both} $k^+$ and in $k^-$.
Therefore, while  $\vec{k}$ has a finite scale \be
 \vec{k} \sim \frac{1}{b} \ee
the longitudinal wavenumbers are negligibly small \be
 k^+,k^- \ll \vec{k} ~. \label{k-hierarchy} \ee

More quantitatively we shall find later that \be
 \frac{k^\pm}{k^i} \sim \frac{\al}{b^{d-3}\, E_{cm}} \ll 1 \label{small-parameter1} \ee
 where $\al$, to be defined later, is schematically the product of
effective charges \be
 \al \sim \hat{q}_R\, \hat{q}_L \ee
 and is the classical and dimensionful analogue of the fine structure
constant. Eq (\ref{small-parameter1}) defines \emph{the hierarchy
of scales and the small parameter of the EFT}. The scattering
deflection angle will be seen to be of the same order as this
small parameter.

The hierarchy of wavenumbers (\ref{k-hierarchy}) is equivalent to
a momentum transfer which is dominantly transverse since \be
 \Delta p_\mu \sim \al\,  k_\mu ~.\ee
Note that unlike quantum field theory the $\al$ factor is
essential on dimensional grounds.


Now we proceed to obtain the field propagator and its correction.
We decompose the kinetic term as follows \be
 4 \pi\, S =  \half \int dx\, \del_\mu \phi ~\del^\mu \phi = -\half \int dx \left|
\vec{\nabla} \phi \right|^2 + \int dx\, \del_+ \phi ~\del_- \phi
\ee Much like the non-relativistic case reviewed above, this
decomposition implies the following Feynman rules for the
perturbation theory: the first and dominant term is considered to
belong to the unperturbed action and determines \emph{the field
propagator} \be
 \parbox[t]{10mm}{\includegraphics[width=10mm]{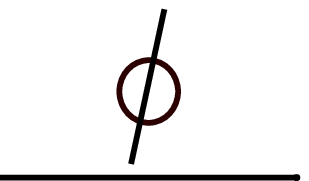}} = 
  \int \frac{d^d k}{(2\pi)^d} \frac{4
 \pi}{\vec{k}^2}   \label{phi-propagator} \ee
 while the second term is subleading and determines a 2-vertex (correction to the propagator) \be
 \parbox{14mm}{\includegraphics[width=14mm]{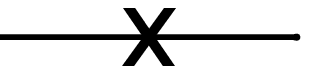}} =
  \frac{k^+ k^-}{4 \pi} \label{phi-prop-corr} \ee

In 4d the transverse space is 2d and we find that in position
space the propagator becomes \bea
 \parbox[t]{10mm}{\includegraphics[width=10mm]{phi-prop.eps}}
 &=&  \int \frac{dk^+\, dk^-\, d^2 k}{(2\pi)^4} \frac{4
 \pi}{\vec{k}^2} \exp \(-i k_\mu \Delta X^\mu\) = \non
 &=& -\log((\Delta \vec{X})^2/\mu_0^2)\, \delta(\Delta z^+)\, \delta(\Delta
 z^-) ~, \label{position-phi-prop}
 \eea
where $\mu_0$ is an arbitrary scale which does not affect the
physics (see appendix \ref{app-prop} for details). From now on we shall set $d=4$.

The transverse propagator (\ref{phi-propagator}) means that within
the EFT the field lines spread instantaneously and only in the
transverse space and thus they are confined to an infinitesimally
thin (due to Lorentz contraction) ``pancake'' shock wave carried
by the particle.

We recall that the transverse propagator applies only to the
potential modes, and will not apply to radiation modes, which are
needed only at higher orders than those calculated in this work.



\presub {\bf The $\phi$-particle vertex}. The particle action
(\ref{particle-action1}) is singular in the massless (and hence
ultra-relativistic) limit $m \to 0$. To overcome that we rescale
$\hat{e} \to m\, \hat{e}$.

In the unperturbed light-like motion of particle R we have
$\vec{x}=z^-=0$ hence it is useful to parameterize the world-line
by $z^+$, namely we choose the gauge \be
 s_R \equiv s^+ = z^+ \label{gauge-choice} \ee
Similarly we choose $s_L \equiv s^- = z^-$ for particle L.

Finally, we perform another rescaling of $\hat{e}$, such that
altogether we define \be
 e:= \frac{p_I}{m} \hat{e} \label{def-e} \ee
 where $p_I$ is the initial light-cone momentum ($I$ stands for
 ``initial''), namely for particle R $p_I \equiv p^+_I$ while for
particle L $p_I \equiv p^-_I$. The $p_I$ factor in the definition
is convenient as it sets the initial $e$ to unity \be
 e_I=1 \label{initial-e} \ee
 as we shall shortly see.

The action (\ref{particle-action1}) becomes \be
 S = p_I \int \( \half \dot{\vec{x}}^2 - \dot{z}^- \) e^{-1} ds - \hat{q} \int \phi(X)\, e ds - \frac{m^2}{2 p_I} \int e\, ds
 ~,
 \label{particle-action2} \ee
 where the effective ultra-relativistic scalar charge is defined by
\be
 \hat{q} := \frac{m}{p_I} q ~. \label{effective-charge} \ee
Note that the $m \to 0$ limit is indeed non-singular now.
 The the 4-momentum is \be
 p^\mu  = p_I\, \frac{dx^\mu}{e\,  ds}
  \label{def-momentum}
\ee
 Computing $p^+_R$  in with the gauge choice (\ref{gauge-choice}) we confirm
that indeed $e_I=1$.

We observe that the new form of action (\ref{particle-action2}) is
very similar to that of a non-relativistic particle: the action is
quadratic in the transverse velocity $\dot{\vec{x}}$ and $p_I$
plays the role of the non-relativistic mass. This is a standard
feature of the light cone or infinite momentum frame
\cite{Dirac:1949cp,Weinberg66,Susskind67}.

The action (\ref{particle-action2}) allows us to read the
$\phi$--particle vertex, namely \be
 \parbox{14mm}{\includegraphics[width=14mm]{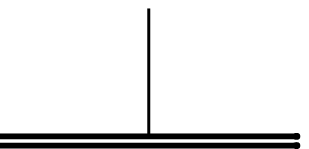}}
 = -\hat{q} \int e\, ds \label{phi-particle-vertex} \ee
 where the double lines represent the particle.


\subsection{Leading scalar, electromagnetic and gravitational momentum transfer}

\noindent {\bf Scalar case}. At first order in the small parameter (\ref{small-parameter1}) one finds
a transverse momentum transfer. In order to compute it we shall need the
$\phi$ propagator (\ref{phi-propagator}), the $\phi$-particle
vertex (\ref{phi-particle-vertex}) and the following equation of motion  \be
 \dot{p}^i = -e\, \hat{q} \del_i \phi \label{p-dot} \ee
 derived from (\ref{particle-action2},\ref{def-momentum}).
 We shall defer the full formulation of the equations of motion and the perturbation theory for the
particle dynamics in the ultra relativistic limit, to the next
section where we compute corrections.

The change in transverse momentum of particle L is opposite to
that of particle R\footnote{As long as we can ignore momentum
leakage to infinity through radiation.} and is called simply ``the
momentum transfer''. The full expression for it is \bea
 \Delta p^i &=&  \hat{q}_L \int e_L ds^- \, (-)\del_i \phi_{R \to L}(X_L) =
 \non
  &=& \hat{q}_R \hat{q}_L \int e_R ds^+\, e_L ds^-\, \del_i G_{L \to R}\(X_R(s^+),X_L(s^-)\)
  \label{full-mom-transfer} \eea
 where $G_{R \to L}(X_R,X_L)$ is the full retarded $\phi$ propagator sourced by $X_R$ at $X_L$ and without
any approximations.

At leading order we may substitute the unperturbed straight line
trajectory for both particles and the leading EFT propagator
(\ref{phi-propagator})  and we have \bea
 \Delta p^{(1)} &=& \hat{q}_R \hat{q}_L \int  ds^+\, ds^-\, \del_i  G^{(0)}_{R \to
 L}\(X_1^{(0)}(s^+),X_L^{(0)}(s^-)\)= \non
 &=& \del_{b^i}  \( \parbox{14mm}{\includegraphics[width=14mm]{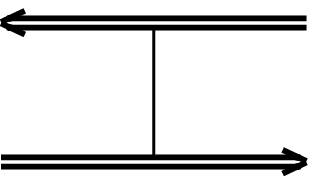}} \) 
 \label{gen-lead-mom-transfer} \eea
This expression describes the momentum transfer in terms of a
diagram which is actually the leading part of the two-body
effective action.

This computation is quite straightforward in the EFT. Using the
position space $\phi$-propagator (\ref{position-phi-prop}) we have
\be
 \parbox{14mm}{\includegraphics[width=14mm]{scalar-1field-exchange.eps}}
 =
 - \int ds^+ ds^- \hat{q}_R \hat{q}_L \log\((\Delta \vec{x})^2 / \mu_0^2\)\, \delta(\Delta z^+)\, \delta(\Delta
 z^-) \label{lead-diag-1} \ee
For light-like particles the integrations annihilate the $\delta$
functions and we have \be
 \parbox{14mm}{\includegraphics[width=14mm]{scalar-1field-exchange.eps}}
 =
 -  \hat{q}_R \hat{q}_L \log\( (\Delta \vec{x})^2/\mu_0^2\) \equiv -\hat{q}_R \hat{q}_L \log(b^2/\mu_0^2)\,  \label{lead-diag-2} \ee
 We found that this result reproduces the computation in the full ``microscopic'' theory.
Substituting back in (\ref{gen-lead-mom-transfer}) we finally
obtain the leading momentum transfer
\be
 \Delta p^{(1)} = -\frac{2 \hat{q}_R \hat{q}_L}{b} ~. \label{lead-scalar-mom-transfer} \ee
The weak scattering angle is can now be stated quantitatively \be
 \Delta \theta =  \frac{\Delta p}{p_z} = \frac{2\, \Delta p}{E_{cm}} = \frac{4 \hat{q}_R \hat{q}_L}{b\, E_{cm}} \label{scatter-angle} \ee
 and it is of first order in our small parameter (\ref{small-parameter1}).

\presub {\bf Electromagnetism}. Now that we reproduced the leading
momentum transfer for a scalar interaction we proceed to generalize
it to the electromagnetic and gravitational interactions. In
electromagnetism the total action is \be
S_{EM,tot} = S[A_\mu] + \sum_{A=R,L} S_A[X_A,\hat{e}_A; \phi] \label{total-EM-action}\ee
where  the field action is \be
 S_{\mbox{Maxwell}} = -\frac{1}{16 \pi} \int d^4X\, \left| F_{\mu\nu} \right|^2 ~, \label{EM-action} \ee
 and the particle action  is \be
 S_p = -\frac{p_I}{2} \int ds ~\dot{X}^2/e -\frac{m^2}{2 p_I} \int e\, ds - q \int A_\mu(X)\, dX^\mu \label{EM-particle-action} \ee
 and where $m$ and $q$ are its mass and electric charge.

Since we assume that the longitudinal gradients are negligible
compared with the transverse ones (\ref{k-hierarchy}) it is
appropriate to perform a dimensional reduction of the
electro-magnetic field onto the transverse plane. Accordingly we
decompose \be
 A_\mu = (\phi_a,A_i) ~~ a=+,-, ~ i=1,2 \ee
 where $\phi_\pm$ are a pair electro-static potentials which are scalars from the transverse perspective.
Since on the unperturbed trajectory $dX^\mu = ds^+ \delta^\mu_+$
for particle R (and similarly for 2) we see that at leading order
particle R couples only to $\phi_+$, while particle L couples only
to $\phi_-$ namely \be
 \sum_{A=1}^2 S_A = -q_R \int \phi_+(s^+)\, ds^+ - q_L \int \phi_-(s^-)\,
 ds^- ~.\ee
This defines the electro-magnetic field-particle vertex, and in
particular there is no need to define effective electric charges
(unlike the effective scalar charges).

The propagator for $\phi_\pm$ in the limit (\ref{k-hierarchy}) can be
read from the $F^2$ kinetic term in (\ref{EM-action}) \be
 S_{EM} \supset \frac{1}{4\pi} \int d^4x\, \vec{\nabla}
 \phi_+\, \vec{\nabla} \phi_- ~. \ee
Note that the sign which is opposite that in the scalar case
signifies a repulsive interaction for identical particles, while
in the scalar case it is attractive.

The longitudinal $SO(1,1)$ boost symmetry of the electro-magnetic
action (\ref{EM-action}) is realized in the EFT as a global
symmetry under which $\phi_\pm$ have charges $\pm 1$. Accordingly we
may represent the fields $\phi_\pm$ by the same \emph{oriented} line
and distinguish them by its orientation, which represents the
flow of charge.

Having determined the Feynman rules we proceed to compute the
leading scattering angle with the same method which we used in the
scalar case (\ref{lead-scalar-mom-transfer}). We find \be
 \Delta p^{(1)} = \del_{b^i}  \mbox{ diagram } = +\frac{2 q_R q_L}{b} \label{lead-EM-mom-transfer} ~. \ee
See \cite{JKO} for a related study which is consistent with ours.

\presub {\bf Gravitation}. We shall now find the field propagator
and field-particle vertex in order to compute the leading
scattering angle for a gravitational interaction.

The gravitational field is \be
 g_{\mu\nu}=\eta_{\mu\nu}+h_{\mu\nu} \ee
  where $\eta_{\mu\nu}$ is the flat space metric and $h_{\mu\nu}$ are gravitational perturbations around it.

The gravitational action is \be
 S_{G} = -\frac{1}{16 \pi G} \int d^4X\, R[g] + \sum_{A=1}^2 S_A \label{grav-action} \ee
 where the particle action is \be
 S_A = -\frac{p_I}{2} \int ds ~\dot{X}^2/e -\frac{m^2}{2 p_I} \int e\,
 ds \ee
 where now \be
 \dot{X}^2 := g_{\mu\nu} \dot{X}^{\mu} \dot{X}^{\nu} = \dot{X}_\mu
 \dot{X}^\mu + h_{\mu\nu} \dot{X}^{\mu} \dot{X}^{\nu} \ee
 and so the field-particle coupling is encoded
by the particle's kinetic term.

Again we perform a dimensional reduction onto the transverse
plane. We see that at leading order particle R couples only to
$h_{++}$, while particle L couples only to $h_{--}$ namely \be
 \sum_{A=1}^2 S_A = -\frac{p^+_{I}}{2} \int h_{++}(s^+)\, ds^+ -\frac{p^-_{I}}{2} \int h_{--}(s^-)\, ds^-  ~.\ee
This defines the gravitational field-particle vertex, where half
the light cone momentum plays the role of an effective
gravitational charge \be
 \hat{q}_{g} = \half p_I ~. \ee

The propagator for $h_{++}, \, h_{--}$ which are both scalars from
the transverse perspective 
 can be read from the Einstein-Hilbert kinetic term in (\ref{grav-action})
\be
 S_{G} \supset -\frac{1}{32 \pi G} \int d^4X\, \vec{\nabla}
 h_{++}\, \vec{\nabla} h_{--} \ee
 (see for example subsection \ref{URG-subsection2}). The negative sign signifies an
 attractive interaction and there is an $8 G$ factor relative to the electromagnetic case. Here the longitudinal boost symmetry
assigns to $h_{++},\, h_{--}$ charges $\pm2$ and again we may
represent the fields by oriented lines.

Altogether the momentum transfer is \be
 \Delta p^{(1)} =  -8G\, \frac{2 \hat{q}_R \hat{q}_L}{b} = -\frac{2 G\, E_{cm}^2}{b} \label{lead-grav-mom-transfer}  \ee
 where we used \be
 E_{cm}^2 = \( p_1^\mu+p_2^\mu\)^2 = 2 p_{1\mu}\, p_2^\mu = 2 p^+_I
 p^-_I ~. \label{Ecm} \ee
This result, eq (\ref{lead-grav-mom-transfer}) reproduces the value in
the literature (see for example eq. (2.7) of \cite{ACV93} as well
as \cite{Dray-`tHooft}).

\presub {\bf Summary of leading momentum transfer}.
Summarizing the three cases above \\
 (\ref{lead-scalar-mom-transfer},\ref{lead-EM-mom-transfer},\ref{lead-grav-mom-transfer}), the leading momentum transfer is \be
 \Delta p^{(1)} = (-)^{\sigma+1}\, \frac{2\,\alpha}{b} \label{momentum-transfer-summary}
\ee
 where $\sigma=0,1,2$ corresponds to the spin of the field and the fine structure constant is given by \be
 \alpha := \left\{ \begin{array}{cc}
 \hat{q}_R \hat{q}_L \equiv \frac{2 m_R m_L\, q_R q_L}{E_{cm}^2} & \mbox{ scalar} \\
 \vspace{.3cm} q_R q_L  & \mbox{ EM} \\
 \vspace{.3cm} G\, E_{cm}^2    &  \mbox{ Gravity}
 \end{array}
  \right. \ee

\section{Two body effective action}
\label{eff-action-section}

In this section we wish to consider higher order corrections
while setting the masses to zero $m_A=0$. Such corrections can arise at one of two steps: either in the two body effective action,
or while solving the equations of motion which arise from it.
As usual the two body effective action is defined by
integrating out the interaction field, namely \be
 S_{2bd} = \mbox{sum of diagrams} \label{2bd-general} \ee
 where one must include all the connected diagrams without quantum loops, namely diagrams which consist of field propagators, field vertices and field-particle vertices, but no particle propagators, such that the diagram is connected and without loops when the source lines are ignored; and external field propagators end on both sources.

In general, and by analogy with the PN approximation we expect 5 possible diagrams (or effects) to  contribute: retardation, vertex corrections, exchange of another field (such as other field polarizations), and finally bulk and world-line
non-linearities.



\presub {\bf Scalar two body potential}. Let us compute the two body
effective potential for the scalar interaction
(\ref{total-scalar-action}).
 Since the field is linear the sum over all diagrams (\ref{2bd-general}) consists only of diagrams with one field exchange between the two sources and may include an arbitrary number of retardation vertices.
Namely \bea
    S_{2bd} &=& \int ds^+ e_R q_R\, \int ds^- e_L q_L G\(X_R(s^+), X_L(s^-) \) = \non
         &=& \alpha \int ds^+\, ds^- e_R e_L \int \frac{d^dk}{(2\pi)^d} e^{i k_\mu (X^\mu_R-X^\mu_L)}\, \frac{4 \pi}{\vec{k}^2}\, R \eea
 where the retardation correction (sum over all retardation vertices) $R$ is given by the expansion \be
    R=\( 1- \frac{2 k^+ k^-}{\vec{k}^2}\)^{-1} = 1 + \frac{2 k^+ k^-}{\vec{k}^2} + \(\frac{2 k^+ k^-}{\vec{k}^2}\)^2  + \dots ~, \label{def-R} \ee
 and \be
    e^{i k_\mu (X^\mu_R-X^\mu_L)} = e^{i\( k^- s^+ - k^+ s^- + k_x b\)}\, T \ee
 where the Taylor coefficient $T$ is given by the expansion \bea
    T &:=& \exp i \(-\vec{k} \cdot \Delta \vec{x} -k^-z^+_L + k^+ z^-_R\) = 1 - i \vec{k} \cdot \Delta \vec{x} +i( k^+ z^-_R -k^- z^+_L) - \half \(\vec{k} \cdot \Delta \vec{x} \)^2 + \dots \non
    \Delta \vec{x} &:=& \vec{x}_R-\vec{x}_L \label{def-T} \eea
Combining all the ingredients we have in the scalar case \be
    S_{2bd,sc} = \alpha  \int ds^+ ds^- \int_{k^\pm} \int_{\vec{k}} e^{i\( k^- s^+ - k^+ s^- + k_x b\)}\, \frac{4 \pi}{\vec{k}^2}\,  T(k^\mu,\Delta X^\mu)\, R\(\frac{2 k^+ k^-}{\vec{k}^2}\)\, e_R e_L \label{2bd-scalar} \ee
where \bea
    \int_{k^\pm} &:=& \int \frac{dk^+ dk^-}{(2 \pi)^2}  \non
    \int_{\vec{k}} &:=&  \frac{d^{d-2}k}{(2\pi)^{d-2}}~. \eea

\presub {\bf Electromagnetism}. Since electromagnetism is a linear as well, we can express $S_{2bd}$ along the lines of the scalar case. First we must complete the discussion of the Maxwell action (\ref{EM-action}). After decomposing $A_\mu \to (\phi_\pm,\, A_i)$  we need to add a gauge fixing term to the action for the transverse vector $A_i$. As usual one could take \be
 S_{GF} = -\frac{1}{8 \pi} \int d^dx  \( \del^i A_i \)^2 \ee
 to render the (transverse) propagators non-singular. However one can still add a $A_i$ independent term to the gauge constraint $\del^i A_i \to \del^i A_i + \dots$. Just like in post-Coulomb approximation (non-relativistic electromagnetism) it is convenient to take the full $d$-dimensional Lorentz gauge thereby avoiding a coupling between $\phi_\pm$ and $A_i$. Accordingly we take \be
 S_{GF} = -\frac{1}{8 \pi} \int d^dx  \( \vec{\nabla} \vec{A} - \del_+ \phi_- - \del_- \phi_+ \)^2 ~. \label{EM-GF} \ee
The total gauge fixed action for the electromagnetic field is \bea
 S_{Max,GF} &=& -\frac{1}{8 \pi} \int d^dx\, \del_\mu A_\nu \del^\mu A^\nu = \\
  &=& -\frac{1}{8 \pi} \int d^dx\,  \left\{ 2 \vec{\nabla} \phi_+ \cdot \vec{\nabla} \phi_- - 4 \del_+\, \phi_+ \del_- \phi_- -\vec{\nabla} A_i \cdot \vec{\nabla} A_i  + 2 \del_+ A_i\, \del_- A_i   \right\} \nonumber \label{EM-action-tot} \eea
The coupling of particle R to the field is \be
 -q_R \int A_\mu dX_R^\mu = -q_R \int ds^+ \( \phi_+ - \vec{A} \cdot \vec{v}_R +\phi_- + \phi_-\, \dot{z}^-  \) \ee
 and similarly for particle L.

The total action above determines the Feynman rules with which we can proceed to calculate the electromagnetic two body action. The sum over diagrams is similar to the scalar case, only now the particles can exchange either $\phi$ with either orientation or $A_i$, which is the magnetic interaction. The factor $e_R\, e_L$ is now absent, and there is an overall sign change, signifying as usual electromagnetic repulsion between like charges. Altogether the two body effective action is the same as in the scalar case (\ref{2bd-scalar}) apart for substituting the following factor \be
S_{2bd,EM} = S_{2bd,sc} ~ /. ~ e_L e_R \to - \( 1 - \vec{v}_R \cdot \vec{v}_L + \dot{z}^-\, \dot{z}^+ \) \label{2bd-EM} \ee

\presub {\bf Ultra Relativistic particle dynamics}. In order to find higher order corrections
for the scattering trajectories we should solve the equations of motion for the corrected two body effective action.
But first we pause to conveniently formulate the equations of motion and the perturbation theory for
 Ultra Relativistic dynamics.

A point-like particle has 3 degrees of freedom. In the
non-relativistic case they can be identified with $\vec{x}(t)$, namely
the spatial coordinates as a function of time. Looking at the
action for particle R (\ref{particle-action2}) we may now identify
the 3 degrees of freedom as follows. $x_\perp$ account for 2 out
of 3.  $z^-$ and $e$ appear in the action only in a single
$s$-derivative term. Hence each requires only a single initial
condition and should be counted as half a degree of freedom.
Together they constitute the remaining degree of freedom.

We find it convenient to use a canonical (Hamiltonian) formalism with respect to the transverse coordinates, but not with respect to $e$ and $z^-$.



\subsection{Second order}

We shall now use the formalism of the two body effective action to go beyond the first order in the small parameter. For concreteness, we specialize to the electromagnetic case and derive the equations of motion from the two body effective action (\ref{2bd-EM}).
 We denote the magnitude of the leading momentum transfer by \be
 \Delta p := \Delta p^{(1)} \equiv \frac{2 \alpha}{b} \ee

First we can recover the first order results including \bea
 x_R^{(1)} &=& -\frac{\Delta p}{p_I^+}\, s^+\, \theta(s^+) \\
 x_L^{(1)} &=& \frac{\Delta p}{p_I^-}\, s^-\, \theta(s^-) \label{xL1} \eea
 where $\theta(s_A)$ is the Heavyside step function.

We proceed to obtain the second order momentum transfer.
By varying the action with respect to $e$ we find \be
 \dot{z}^- = \half \dot{\vec{x}}^2 \ee
 from which we have \be
 \Delta p^-_R = \half \frac{(\Delta p)^2}{p^+_I} ~. \label{Delta-p-} \ee

Varying the action with respect to $z^-$
we find that in order to saturate the $k^+$ factor we should extract an $s^-$ factor by expanding $T=1 + i \vec{k} \vec{x}_L + \dots$ where $\vec{x}_L$ is given by (\ref{xL1}).
Integrating over $ds^+$ we obtain the total momentum change \be
 \Delta p^+_R = -\half \frac{(\Delta p)^2}{p^-_I} \label{Delta-p+} \ee

The total transverse momentum change can be seen to vanish at this order.

Our results for the second order momentum change are related to the first order results in the following way. The change in $p^-$ (\ref{Delta-p-}) is required in order to preserve the mass-shell condition $p^2=2 p^+ p^- - \vec{p}^2=0$. The change in $p^+$ is such that in the center of mass frame where $p^+_I=p^-_I=E_{cm}/2$ the change in energy (of each particle) vanishes $\Delta E = \( \Delta p^+ + \Delta p^-\)/\sqrt{2} =0$ as one could expect on general grounds. Actually (\ref{Delta-p+}) is the unique generalization outside the center of mass frame as it has the correct transformation law under a longitudinal boost.


\section{Interaction duration at the third order}
\label{interaction-time-section}

In this section we compute the interaction duration thereby resolving the instantaneous approximation. For that purpose we introduce an improved ``renormalization condition''.

\presub {\bf An improved renormalization condition}. The perturbative computations become simpler if one takes the initial conditions not at early time $t=-\infty$ but rather at the interaction time $t=0$, since the latter enjoys a symmetry, namely time reversal, at least as long as radiation is ignored. For that purpose one pretends that the system's state at $t=0$ is known and propagates both forward and backward in time, namely $s_{\pm \infty}=s_{\pm \infty}(s_0)$ where $s_t$ is a provisional notation for the system's state at time $t$ (actually the two relations are essentially the same due to the symmetry). Then one solves for $s_0=s_0(s_{-\infty})$ and substitutes back into $s_{+\infty}=s_{+\infty}(s_0)$ to obtain the required $s_{+\infty}=s_{+\infty}(s_{-\infty})$. By way of analogy we refer to this procedure as a change of renormalization condition.

We demonstrate this procedure at the leading order. Using early time initial conditions we already found the leading momentum transfer to be (\ref{momentum-transfer-summary}) where $b$ is the standard impact parameter at early time. Repeating the calculation with initial conditions at $t=0$ we obtain the same result only with $b \to b_0$ where $b_0$ is the distance between the two bodies at $t=0$.  The scattering angle $\theta$ (in the center of mass frame) is given by \be
 \sin \frac{\theta}{2} =\frac{\alpha}{b_0\, p_0} \ee
  where $p_0=\left| \vec{p} \right|$ (it is the same for both bodies), and $b_0$ is related to $b_i$ via \be
  \cos \frac{\theta}{2} =\frac{b}{b_0} \ee
 altogether yielding \be
 \tan \frac{\theta}{2} = \frac{\alpha}{b\, p_0} ~.\ee
This last expression sums beyond leading contributions to $\theta$ from expanding the left hand side, yet additional higher order contributions could arise on the right hand side as well.

\presub {\bf Interaction duration}. Repeating the perturbative procedure with $t=0$ initial conditions the first order results (\ref{xL1}) change slightly into \be
 x_L^{(1)} = -\frac{\Delta p}{p_0}\, s^-\, \theta_a(s^-) \ee
 where $\theta_a(s):= \theta(s)-\half \equiv \mbox{sgn} (s)/2$. The antisymmetry of the step function $\theta_a$ causes certain cancelations at higher orders which were not present in the original perturbation theory thereby fulfilling our expectation for simplifications. In particular it allows us to proceed to compute the transverse accelerations up to third order. The full computation is presented in appendix \ref{duration-comp-app}. We obtain \be
 p_0^+\, \ddot{x}_R = -\frac{2\al}{b}\, \delta(s_+) -\al \(\frac{\Delta p}{2 p_0}\)^2 b\, \delta''(s^+) \label{third-order-x-ddot} ~.\ee

We note that $\ddot{x}^{(3)}$ is proportional to $\delta''(s^+)$. First, this means that the momentum transfer vanishes at this order. Moreover, since $\ddot{x}^{(1)} \propto \delta(s^+)$, the ratio of the corresponding prefactors has dimensions of time squared and can be taken as an estimate for the (square) of the interaction duration $\tau$. More precisely, a function of the form \be
 const\, \(\delta(s) + \half \tau^2 \delta'' (s)\, \) \ee
 implies that $\tau$ is (the standard deviation of) its duration, and in particular it is finite, despite the use of delta functions.
 Following (\ref{third-order-x-ddot}) we find  \be
 \tau = \frac{\Delta p}{2 p_0} b = \sin \frac{\theta}{2}\, b = \frac{\al}{p_0} \label{interaction-time} ~. \ee
 We pause to make several comments on this result.
  First, it is important that the relative sign between the terms in (\ref{third-order-x-ddot}) is positive, allowing us to have a positive $\tau^2$, while the opposite would mean that counter-intuitively $\ddot{x}(s)$ does not have a constant sign.
  Secondly, we note that $\tau/b$  is the small parameter of the EFT (\ref{small-parameter1}) and hence the EFT's basic tenet for a hierarchy of scales can now be expressed as $\tau \ll b$, namely the longitudinal interaction length is much smaller than the transverse impact parameter.
 Thirdly, the constants in (\ref{interaction-time}) have a simple physical interpretation: a deviation in direction through an angle $\theta$ over a  duration (arc-length) $\tau$ implies that the radius of curvature is $b$, namely the center of curvature for the motion of particle R resides, as expected, at particle L which exerts the force on it.

We note that the result (\ref{third-order-x-ddot}) provides a strong test for our EFT: it contains integrals which diverge for $d=4,6,8$ (poles in dimensional regularization) yet the final answer is finite for all $d \ge 4$. In particular, the cancellation of the pole at $d=4$ relies on the sum of three different terms. Moreover, the sign and the precise numerical factor of the third order part match exactly with expectation, as explained.

\section{Non-zero mass corrections}
\label{mass-corr-section}


We now turn to compute non-zero mass corrections in the scalar case.\footnote{ At this order it would be essentially the same also for EM and gravitational fields.}  Here one only needs to make
a relatively small correction in the derivation of the leading
momentum transfer. When passing from (\ref{lead-diag-1}) to
(\ref{lead-diag-2}) one must retain a Jacobian from the
integration over $\delta$ functions as follows.
\be
 \int ds^+ ds^- ~ \delta (\Delta z^+)\, \delta (\Delta
 z^-) = \int ds^+ ds^- ~ \delta^{(2)}(s^+,s^-)\, \(
 \frac{\del(\Delta z^+,\Delta z^-)}{\del(s^+,s^-)} \)^{-1} = \(
 \frac{\del(\Delta z^+,\Delta z^-)}{\del(s^+,s^-)} \)^{-1}
 \ee

To evaluate the Jacobian we note that \bea
 \Delta z^+ &=&  s^+ - \frac{p_L^+}{p_L^-}\, s^-
 \non
 \Delta z^- &=&  -\frac{p_R^-}{p_R^+}\, s^+ + s^- \label{z-n-s}
 \eea
Neglecting the transverse motion (more precisely $\vec{p}^2 \ll
m^2$) we have \be
 \frac{p^-_R}{p^+_R} = \frac{m_R^2}{2 (p_R^+)^2} = e^{-2 \theta_R} \label{jacob2} \ee
 where initially (and approximately throughout the motion) $p_1^+ \simeq p^+_I$, $\theta_R$ is the rapidity ($\theta_R := \mbox{arctanh }
\beta_R$),  and similarly for particle L. From
(\ref{z-n-s},\ref{jacob2}) we compute the Jacobian \be
   \( \frac{\del(\Delta z^+,\Delta z^-)}{\del(s^+,s^-)} \) =
   1-e^{-2\theta} = 1 - \frac{m_R^2\, m_L^2}{4 (p^+_I)^2\, (p^-_I)^2} = 1-\frac{m_R^2\, m_L^2}{E_{cm}^4} \label{Jacobian} \ee
 where the relative rapidity is $\theta:=\theta_R+\theta_L$, and in the last equality we used (\ref{Ecm}).

This Jacobian corrects both the effective action diagram
(\ref{lead-diag-2}) and the momentum transfer
(\ref{lead-scalar-mom-transfer}) which becomes \be
 \Delta p^{(1)} = -\frac{2 \hat{q}_R \hat{q}_L}{b} \( 1-\frac{m_R^2\, m_L^2}{E_{cm}^4} \)^{-1} ~.
 \label{m-corrected-scalar-mom-transfer} \ee
This expression could be expanded as a power series in the following small
parameter \be 
 \frac{m_R^2\, m_L^2}{E_{cm}^4} \ll 1
\label{small-param2} ~.\ee

\section{The Fields of Ultra-Relativistic Gravitation}
\label{URG-section}

In this section we treat the case of the gravitational interaction in more detail.
In the limit of weak ultra-relativistic scattering the components of the gravitational field split into
several fields from the transverse perspective. In the literature
(see for example \cite{ACV93}) the gravitational Einstein-Hilbert
action is expanded perturbatively for weak gravitational fields.
Here we shall define the field decomposition
(\ref{decomposition}) and compute its whole non-linear action
(\ref{URG-action}). Expanding it allows to readily read off all
the gravitational bulk propagators and vertices for the
perturbation theory. For example we reproduce a leading propagator
term in (\ref{Hprop}).

In the related case of the post-Newtonian approximation it is natural and useful to decompose the gravitational
field through a temporal Kaluza-Klein reduction into
Non-Relativistic Gravitational (NRG) fields
\cite{CLEFT-caged,NRG}. These consist of the Newtonian potential,
the gravito-magnetic 3-vector and a spatial metric. The full,
non-linear gravitational action for these fields was determined in
\cite{NRGaction}.

Here we go further by allowing dimensions larger than one (and
arbitrary in principle) for both (the longitudinal) fiber and (the
transverse) base. Computing the action by the standard method metric $\to$
Christoffel symbols $\to$ curvature tensor $\to$ action would be a
very complicated analytical task, perhaps hopelessly so. Here we
simplify the computation to manageable form with no computerized
computation by using a non-orthonormal frame within Cartan's
method, namely a hybrid method which incorporates both a
non-trivial frame and a non-trivial metric as in \cite{NRGaction}.\footnote{I was notified that an action whose mathematical form is essentially the same was given by Yoon \cite{Yoon}.
 While the mathematical context and tools there are essentially the same as here, namely a Kaluza-Klein reduction, the physical context and application is completely different and has no relation to ultra-relativistic gravitation. Instead \cite{Yoon} interprets 4d General Relativity as a 1+1 gauge theory. Technically, there the $1+1$ space is the base while here it is a fiber, and here spacetime dimension is arbitrary.
 I thank S. Carlip for alerting me to \cite{Yoon}.}

\presub {\bf Field decomposition and action}.
We work in the center of mass frame of a $d$ dimensional spacetime
and we denote the longitudinal direction by $z$, and the
transverse directions by $x^i$.

In the leading ultra-relativistic limit transverse gradients
dominate over longitudinal ones \be
 \del_a \equiv \del_{z\pm} \ll \del_i \equiv \del_x \ee
Therefore it is natural to perform a dimensional reduction \`{a}
la Kaluza-Klein (KK) \cite{Kaluza-Klein} of the metric over the
light-cone coordinates $z^\pm$, a reduction which highlights the
transformation properties (or tensor nature) with respect to gauge
transformations which depend only on the transverse directions $x$
\bea
 g_{\mu\nu} &\to& \(G_{ab},\, A^a_i,\, g_{ij}\) \non
 ds^2 &=& G_{ab} ( dz^a-A^a_i dx^i ) ( dz^b-A^b_j dx^j ) - g_{ij} dx^i dx^j  \label{decomposition}
 \eea
 In this expression $a,b=+,-$ and $i,j=1,2,\dots,d-2$. Note that
 this dimensional reduction is more general and it applies to a
reduction to a any base manifold $X$ parameterized by $x^i$ over a
any fiber $Z$ (not necessarily 2d) parameterized by the
coordinates $z^a$. We observe that from the transverse perspective
the fields are $G_{ab}$ a symmetric matrix of 3 scalars, $A^a_i$ a
pair of transverse vectors, and $g_{ij}$ the transverse metric. To
put this dimensional reduction in the context of the literature we
recall that the original and standard KK reduction
\cite{Kaluza-Klein} is over spatial directions, the
Non-Relativistic Gravitational reduction (NRG)
\cite{CLEFT-caged,NRG} is over the time direction, while here the
reduction is over a Lorentzian $1+1$ fiber.

The action is simply \bea
 S &=& \frac{1}{16 \pi G} \int \sqrt{-G} \sqrt{g}
d^{d-2}x d^2z  \cdot \non & &
 \left\{- \left<
K_{abi}[G]\right>_{dW}^2  + \frac{1}{4} \bar{F}^2 + \bar{R}[g]
\right. 
 \left. +\left< \half \del_a g_{ij} \right>_{dW}^2 - R[G] \right\}
 \label{URG-action}
\eea

We proceed to define all the symbols and conventions. First \bea
 G &:=& \det G_{ab} \non
 g &:=& \det g_{ij} \\
 D_i &=& \del_i  + A^a_i \del_a
 \eea

The extrinsic curvature of the $1+1$ fiber is
 \footnote{In general the extrinsic curvature $K$ evaluated on two vector fields $X,Y$ which lie in a sub-manifold
 is defined by $K(X,Y)=\(D_X Y \)^\perp$. This defines a symmetric tensor.}
\be
 K_{abi}[G] := - \half \( D_i G_{ab}+ G_{c(a} \del_{b)} A^c_i \) = - \half \( \del_i G_{ab}+ {\cal L}_{A_i}
 G_{ab} \)
 \ee
 where in the last expression ${\cal L}_{A_i}$ denotes the Lie
derivative with respect to the longitudinal vector $A_i \equiv
A_i^a$.
 \footnote{Recall that the Lie derivative by a vector field
$V$ of a vector field $W$ is defined as ${\cal L}_V W^\mu := \[
V,W \] := V^\nu \del_\nu W^\mu - W^\nu \del_\nu V^\mu$ where $\[
V,W \]$ denotes the commutator of the two vector fields. When one
extends this derivation to all tensors one finds that the Lie
derivative of a co-vector $\omega$ is given by ${\cal L}_V
\omega_\mu = V^\nu \del_\nu \omega_\mu + \omega^\nu \del_\mu
V^\nu$. Similarly ${\cal L}_V G_{ab} = V^c \del_c G_{ab} + G_{ac}
\del_b V^c + G_{bc} \del_a V^c$.}
 The $(-1/2)$ prefactor was inserted to conform with the
standard definition of the extrinsic curvature.

For any symmetric tensor field we define a ``deWitt'' quadratic
form (actually once applied to a differential of a metric and
integrated over the manifold it becomes a metric on the space of
metrics \cite{deWitt-metric}, see \cite{NRGaction} for its
appearance in the NRG action) \be
 \left< h_{IJ}\right>_{dW}^2 := \left| h_{IJ} \right|^2 - h^2
 \ee
 In particular \bea
 \left< K_{abi}[G]\right>_{dW}^2 &=&   g^{ij} G^{ac} G^{bd}\, \( K_{abi} K_{cdj} - K_{aci} K_{bdj} \) \non
 \left< \half \del_a g_{ij} \right>_{dW}^2 &=& \frac{1}{4} G^{ab}\, g^{ik} g^{jl} \[  \del_a g_{ij}\, \del_b g_{kl} - \del_a g_{ik}\, \del_b g_{jl} \] \eea

The generalized (magnetic) field strength\footnote{Note that as
usual the $1/4$ prefactor could have been avoided had we
accompanied the anti-symmetrization in the definition of $\bar{F}$
with a division by $2$.}
 is defined by
\be
  \bar{F}_{ij}^a := D_i A_j^a - D_j A_i^a = F_{ij}^a + A_i^b \del_b A_j^a -
  A_j^b \del_b A_i^a \label{def-barF}
\ee
 and its square is given by \be
\bar{F}^2 = G_{ab} g^{ik} g^{jl}\, F_{ij}^a F_{kl}^b ~.
 \ee

Finally $\bar{R}[g]$ denotes the Ricci scalar of the transverse
metric $g$ where the derivatives in its expression are replaced
everywhere as follows $\del_i \to D_i$. Borrowing notation from
the {\it Mathematica} software this definition can be stated by
\be
 \bar{R}[g] := R[g] ~ /. ~ \del_i \to D_i ~.\ee

\presub {\bf Derivation}. In order to compute the action we used a
non-orthonormal frame within Cartan's method, namely a hybrid
method which incorporates both a non-trivial frame and a
non-trivial metric as in \cite{NRGaction}. This action generalizes
an analogous result from the KK literature found by Aulakh and
Sahdev \cite{AulakhSahdev}, and the NRG (Non-Relativistic
Gravitational) action \cite{NRGaction}.\setcounter{footnote}{10} \footnotemark[\value{footnote}] \setcounter{footnote}{13} 

\presub {\bf Tests}. We tested the Ultra-Relativistic
Gravitational (URG) action (\ref{URG-action}) in several limits.
For $d_Z=1$ we reproduce the KK \cite{AulakhSahdev} and NRG
actions.
For $d_X=1$ we reproduce the ADM \cite{ADM} action (see for
example \cite{NRG-ADM}).
As a Final test for $A^a_i=0$ the action is symmetric with respect
to the exchange $X \leftrightarrow Z$.
The ``stationary limit'', namely no $z$ dependence, is another
interesting limit. In this limit the last two terms in the action
(\ref{URG-action}) vanish,\footnote{If a curved $Z$ fiber is
allowed then some $R[G]$ would remain.} while the other terms
simplify: $D_i \to \del_i, ~ \bar{R} \to R$, $\bar{F} \to F$ and
$K_{abi} \to -(1/2) \del_i G_{ab}$.

\subsection{Perturbing around flat spacetime}
\label{URG-subsection2}

The dimensionally reduced action can be used to linearize the
action around any prescribed product space-time $X \times Z$. In
our case the unperturbed space-time is flat and accordingly we may
use \bea
 G_{ab} &=& \eta_{ab} + H_{ab} \non
 g_{ij} &=& \delta_{ij} + h_{ij} ~. \eea

At leading ultra-relativistic order one source couple dominantly
to $H^{++}$ while the others couples to $H^{--}$. For 2d metrics
the deWitt metric simplifies \be
 \left< G_{ab} \right>^2_{dW} = \frac{2}{-G}  \( dH_{++} dH_{--} - dH_0^2 \)\non
\ee
 where \bea
 H_0 &:=& H_{+-} \non
 -G &=& (1+H_0)^2-H_{++} H_{--} ~.
 \eea
Substituting into (\ref{URG-action}) we find that the propagator
for $H_{++}, \, H_{--}$ is  \be
 S \supset -\frac{1}{32 \pi G} \int d^{d-2}x\, d^2z\, \vec{\nabla}_\perp
 H_{++}\, \vec{\nabla}_\perp H_{--} ~. \label{Hprop} \ee

Longitudinal boosts are a global $SO(1,1) \simeq \IR$
 symmetry of the action.\footnote{From the
transverse point of view $G_{ab}$ are three scalars and being a
quadratic form the action is invariant under a similarity
transformation $G \to R^T G R$ for any $R \in GL(2,\IR)$. Moreover
the vacuum (unperturbed solution) $G_{ab}=\eta_{ab}$ is invariant
under the $SO(1,1) \simeq \IR$ subgroup of 2d Lorentz
transformations, which accordingly is a symmetry of the linearized
action.} Under this symmetry $H_{++}$ has charge $+2$ while
$H_{--}$ has charge $(-2)$. This symmetry can be represented in
the Feynman rules by representing the fields $H_{++},\, H_{--}$ by
an oriented line and distinguishing them by its orientation, which
represents the flow of charge.

\section{Discussion}
\label{discussion-section}


In this paper we set-up a classical effective field theory for weak ultra-relativistic scattering. In this limit the field lines are confined to the transverse directions and they spread instantaneously. These properties are captured by the propagator $1/k_\perp^2$. The instantaneous nature of the fields is similar to the non-relativistic limit, only in one fewer dimension.

The three degrees of freedom of a particle consist of two transverse degrees of freedom which are of non-relativistic nature and a pair of longitudinal half degrees of freedom.

The small parameter (\ref{small-parameter1}) is the ratio of leading longitudinal momentum transfer to the transverse one, and it is of the same order as the scattering angle (\ref{scatter-angle}).

To establish power counting we decompose the quadratic action to a
leading part and corrections. The particle -- field coupling is
decomposed (for non-scalar fields) through a dimensional reduction
onto the transverse plane. The bulk action of the fields is
decomposed accordingly.

We applied the EFT to reproduce the leading ${\cal O}\(\alpha\)$ transverse momentum transfer for the various interaction fields (\ref{momentum-transfer-summary}). At order ${\cal O}\(\alpha^2\)$ we computed the leading longitudinal momentum transfer (\ref{Delta-p-},\ref{Delta-p+}). At order ${\cal O}\(\alpha^3\)$ we computed a certain subleading corrections to the transverse momentum transfer including a retardation effect. In addition we computed corrections due to another small parameter: a non-zero mass.

We conclude with a discussion of \emph{open questions}.

\presub {\bf Radiation}. Just like in PN at some order we must
take into account radiation effects. What is the typical radiation
frequency? At order ${\cal O}\(\alpha\)$ the momentum change is
sudden, and hence there is no frequency scale. Our third order result for the interaction time (\ref{interaction-time}) suggests a scale for
 the radiation frequency, namely $k_{rad} \sim \tau^{-1} \sim \( \eps b\)^{-1}$.
 Radiation will break the conservation of energy,
momentum and angular momentum, and it would be interesting to
compute the leading violation rates. In reaction, the particle
will feel a radiation reaction force.

\presub {\bf Quantum theory}. Our effective theory could be
promoted to a quantum theory simply by allowing loops. At leading
order the quantum theory is controlled by the classical one
through the eikonal approximation (see for example
\cite{eikonal}). However, the author does not know whether the
classical theory is available or was it a missing link so far. It
remains to be seen whether obstructions arise at loops.

\presub {\bf Higher $\al$ orders}. An obvious extension is to
compute higher order effects. First, one could compute higher
non-linear terms in the two body effective action, for instance in
gravity. Secondly, one could solve the equations of motion (which
include more information than the momentum transfer). Here the
differential equations of motion degenerate to the instant of
passing and hence become algebraic in nature, and it might
be possible to obtain explicit expressions.

\presub {\bf Finite size effects}. Finally, one may go beyond the
particle approximation and consider finite
 size effects, including spin and charge  multipole moments.

\presub {\bf Ultra-relativistic gravitation}. We note two issues. First, in 4d spacetime the transverse metric
is 2d which may bring about additional simplifications.
Secondly, in the post-Newtonian case Weyl rescaling of the metric was employed, and actually was necessary even to reproduce the Newtonian potential.
In 4d space the transverse space is 2d and Weyl rescaling is less
effective, but it could possibly be of use at least in higher
dimensions.

\subsection*{Acknowledgments}

I thank Walter Goldberger for collaboration during early stages of
this work including during the setting up of the effective field
theory, and I thank Gabriele Veneziano for stimulating
correspondence.

\appendix

\section{Light cone conventions}
 \label{app-conventions}

We define the transformation to light cone coordinates along the
$z \equiv x^{d-1}$ axis as follows
 \be z^\pm = \( t \pm z \) /\sqrt{2} \label{def-light-cone} \ee
 Accordingly the components of any spacetime vector $v^\mu, ~\mu=0,\dots,d-1$ are
\be
 v^\mu \equiv (v^+,v^-,\vec{v}) \ee
where arrowed vectors in this paper denote transverse ($d-2$)
vectors \be
 \vec{v} \equiv v_\perp \equiv v^i ~ i=1,\dots,d-2 ~.\ee

The definition (\ref{def-light-cone}) implies \bea
 x \cdot y &=& x^+ y^- + x^- y^+ - \vec{x} \cdot \vec{y} \non
 d^d x &=& dx^+ dx^- d^{d-2} x \non
 \eea
We shall only need these latter relations so the $\sqrt{2}$ factors
 shall never appear in final results.

\section{EFT Propagator in position space}
\label{app-prop}

We wish to compute the EFT propagator (\ref{phi-propagator}) in
position space 4d. It is given by the Fourier transform \be
 \phi \mbox{ prop } =  \int \frac{dk^+\, dk^-\, d^2 k_\perp}{(2\pi)^4} \frac{4
 \pi}{k_\perp^2} \exp \(-i k_\mu \Delta X^\mu\) ~. \ee
The $k_\perp$ integral is evaluated through application of the
formula \be
 \int \frac{d^{\hat{d}}k}{(2 \pi)^{\hat{d}}} \, \frac{1}{(k^2)^\alpha}\, \exp(i k
 \cdot x) = \frac{1}{(4 \pi)^{\hat{d}/2}}\,
 \frac{\Gamma(\frac{\hat{d}}{2}-\alpha)}{\Gamma(\alpha)}
 \(\frac{x^2}{4} \) ^{\alpha-\hat{d}/2} \label{integ-formula1} \ee
In 4d we use dimensional regularization and take the transverse
dimension to be $\hat{d}=2-2 \eps$ and we obtain (in the
$\bar{MS}$ scheme) \be
 \int \frac{d^dk_\perp}{(2 \pi)^d} \, \frac{1}{(k_\perp^2)}\, \exp(i k_\perp \cdot \Delta x_\perp) \to
-\frac{1} {4 \pi} \log\(\frac{r_{12}^2}{4} \) ~.\ee

Substituting back we have \be
 \phi \mbox{ prop } = -\log(\Delta x_\perp^2)\, \delta(\Delta z^+)\, \delta(\Delta  z^-) ~. \ee

\section{Computation of interaction duration}
\label{duration-comp-app}

In this appendix I present the detailed calculation of the third order term (\ref{third-order-x-ddot}) which resolves the duration of the interaction.

The total action is \be
 S = \sum_{A=R,L} S_A + S_{2bd} \ee
 where the free action for the R particle is \be
 S_R = p_0^+ \int ds^+ \( \half \dot{\vec{x}}^2 - \dot{z}^- \)/e_R \ee
(with an analogous one for particle L) and the two-body effective action specialized to the case of the electro-magnetic interaction
 is taken from (\ref{2bd-scalar},\ref{2bd-EM}) \be
 S_{2bd,EM} = -\al \int_{s^\pm\, k^\pm\, \vec{k}} \frac{4 \pi}{\vec{k}^2} E(s^\pm,k^\mu,b)\, T(k^\mu,\Delta X^\mu)\, R\(\frac{2 k^+ k^-}{\vec{k}^2}\)\, M(\dot{X}_A^\mu) \ee
 where $R,T$ are defined by (\ref{def-R},\ref{def-T}) and \bea
 E &:=& \exp i\(k^- s^+-k^+ s^- + k_x b \)      \non
 M &:=&  1 - \vec{v}_R \cdot \vec{v}_L + \dot{z}_R^-\, \dot{z}_L^+  ~. \eea

By expanding the equations of motion we proceed to compute the transverse acceleration $\ddot{x}_R$ at order $\al^3$. All the following equalities hold up to terms from other orders. \bea
 & & \al^{-1}\, p_0^+\, \ddot{x}_R = \non
  &=&  \al^{-1} \frac{\delta}{\delta x_R} S_{2bd} = \non
    &=& \int_{s^-\, k^\pm\, \vec{k}} \frac{4 \pi}{\vec{k}^2} \del_b E\, T\, R\, M - \frac{d}{ds^+} \int \frac{4 \pi}{\vec{k}^2} E\, T\, R\, \dot{x}_L = \non
    &=& \int \frac{4 \pi}{\vec{k}^2} \del_b E\, R\, \[-\half \(\vec{k} \cdot \Delta\vec{x}^{(1)}\)^2 + i \(k^+ z_R^{-(2)} - k^- z_L^{+(2)}\) - \dot{x}^{(1)}_R \dot{x}^{(1)}_L)  \]  \non
   & &   ~~~~~ -\frac{d}{ds^+} \int \frac{4 \pi}{\vec{k}^2} E\,  R\, \dot{x}_L^{(1)} \(-i \vec{k} \cdot \Delta\vec{x}^{(1)} \) = \non
    &=&  \int \frac{4 \pi}{\vec{k}^2} \del_b E\, R\, \(-\half k_x^2 x^{(1)}_L\, ^2 - i k^- z_L^{+(2)} \) - \int \frac{4 \pi}{\vec{k}^2} i k^-\, \del_b E\,  R\, \dot{x}_L^{(1)} x_L^{(1)} = \non
   &=&  \int \frac{4 \pi}{\vec{k}^2} \del_b E\, R\, \( -\half k_x^2\, s^-\,^2 \dot{x}_L^{(1)}\,^2 -\half i k^- s^- \dot{x}_L^{(1)}\,^2 - i k^- \dot{x}_L^{(1)}\, x_L^{(1)} \) = \non
    &=& \(\frac{\Delta p}{2 p_0}\)^2 \int \frac{4 \pi}{\vec{k}^2} \del_b E  \[ -\half k_x^2\, s^-\,^2 \(\frac{2 k^+ k^-}{\vec{k}^2}\)^2
        -\frac{3}{2}i k^- s^- \frac{2 k^+ k^-}{\vec{k}^2} \] = \non
    &=& -\(\frac{\Delta p}{2 p_0}\)^2\, \delta''(s^+) \int_{\vec{k}} \frac{4 \pi}{\vec{k}^2} \del_b E \[ -2 k_x^2 \frac{(-\del_-^2 s^-\, ^2)}{(\vec{k}^2)^2} -3i \frac{(-i\del_- s^-)}{\vec{k}^2} \] = \non
    &=& \(\frac{\Delta p}{2 p_0}\)^2\, \delta''(s^+) \[ 4 \del_b^3 \int \frac{4 \pi}{(\vec{k}^2)^3}\, E + 3 \del_b \int \frac{4 \pi}{(\vec{k}^2)^2}\, E \] = \non
    &=&  \(\frac{\Delta p}{2 p_0}\)^2\, \delta''(s^+) \[ 4 \del_b^3 \frac{1}{(4\pi)^{\hd/2-1}}\, \frac{\Gamma(\frac{\hd}{2}-3)}{2} \(\frac{b}{2}\)^{6-\hd}
    +3 \del_b \frac{1}{(4\pi)^{\hd/2-1}}\, \Gamma(\frac{\hd}{2}-2)\, \(\frac{b}{2}\)^{4-\hd} \] = \non
    &=& \(\frac{\Delta p}{2 p_0}\)^2\, \delta''(s^+)\, \frac{1}{(4\pi)^{\hd/2-1}}\, \(2 \frac{5-\hd}{2} -3 \) \Gamma(\frac{\hd}{2}-1)\, \(\frac{b}{2}\)^{3-\hd} = \non
    &=& -\(\frac{\Delta p}{2 p_0}\)^2\, \delta''(s^+)\,b ~ \frac{1}{(4\pi)^{\hd/2-1}}\, \Gamma(\frac{\hd}{2})\, \(\frac{b}{2}\)^{2-\hd} \to \non
    &\to& -\(\frac{\Delta p}{2 p_0}\)^2\, \delta''(s^+)\,b \eea
 where each new line was gotten as follows
\begin{enumerate}
 \item The action was varied while using that $e_R=1$ up to this order.
 \item In our coordinate system (see figure \ref{Geometry})) $\frac{\del}{\del x_R} = -\del_b$.
 \item We expand $T$ to retain all terms of the requested order.
 \item Only terms of weight 2 with respect to $k^-,s^-$ will survive.
 \item Substituting \bea
  x_L^{(1)} &=& s^-\, \frac{\Delta p}{2 p_0} \non
  z_L^{+(2)} &=& s^- \dot{z}_L^{+(2)} =\half s^- \dot{x}_L^{(1)}\,^2 \eea
 \item Expanding $S$ such as to saturate each factor of $s^-$ by the same factor of $k^+$ -- other terms do not survive. The two last terms in the parenthesis are of the same form and their coefficient sum as follows $1/2 + 1 = 3/2$.
 \item Integrating over $s^-,k^\pm$ by replacing  $k^-\, ^2\to -\del_+^2$ and  $k^+\, \to -i\del_-$.
 \item Replacing $k_x^2 \to -\del_b^2$.
 \item Performing the $\vec{k}$ integration by using (\ref{integ-formula1}). We define $\hd := d-2$ to be the transverse dimension.
 \item Performing the $\del_b$ derivatives allows to cancel the original poles at $d=6,8$ but leaving the one at $d=4$ (namely $\hd=2$).
 \item The two terms in the parenthesis combine exactly to $2-\hd$ thereby cancelling the Gamma function pole at $d=4$.
 \item Finally we take the limit $d \to 4$.
\end{enumerate}

\section{The ring $\IR^{1,1}$}

The scalar doublet of fields $\phi_\pm$ carry charges $\pm 1$. They share many properties with the complex (real doublet) scalar charged field.
In particular the kinetic term $S \supset 2 \del \phi_+ \del  \bar{\phi_-}$ is analogous to the standard complex kinetic term $S \supset \del \phi \del  \bar{\phi}$, and the functional variations $\delta_+\equiv \delta/\delta \phi_+, ~\delta_- \equiv \delta/\delta \phi_-$ are analogous to the $\delta, ~\bar{\delta}$ variation operators.
    This analogies can be traced back to an interesting algebraic structure on the 2d Minkowski space $\IR^{1,1}$ which we now proceed to outline.

A point $z$ in  $\IR^{1,1}$ is specified by a pair \be
 z \equiv (z^+,z^-) ~. \ee
  Addition is defined as usual by $z+w:=(z^+ + w^+,z^- + w^-)$. In analogy with the complex case we wish to define an inner product $\left\| \right\|^2$ with real values, only here it is not positive definite and it is given by the metric \be
  \left\| z \right\|^2 = 2\, z^+\,  z^- \ee
We wish to define a multiplication which respects the inner product, namely \be
 \left\| z \cdot w \right\|^2 = \left\| z \right\|^2  \left\|  w \right\|^2 \ee
This can be achieved by defining multiplication through \be
 z \cdot w:=(\sqrt{2}\, z^+\, w^+, \sqrt{2}\, z^-\, w^-) \ee
This multiplication defines a commutative ring with unit $1 \equiv (1,1)/\sqrt{2}$.
Actually this definition is not unique, for any $r \neq 0 \in \IR$ we could take $z \cdot w:=(\sqrt{2}r\, z^+\, w^+, \sqrt{2}\, z^-\, w^-/r)$ and accordingly the value of the unit would change along the curve $ \left\| z \right\|^2=1$.

Another point of analogy with the complex numbers is the existence of a conjugation which is a ring automorphism, given by \be
 \bar{z} = (z^-,z^+) \ee

There are of course certain differences between $\IR^{1,1}$ and $\IC$. First, $\IR^{1,1}$ is not a field, since $(r,0),\,(0,r) ~\forall r \neq 0 \in \IR$ has no inverse. As another difference we note that the fundamental theorem of algebra does not hold here. For instance, the equation $z \cdot z =1$ has four solutions $z=(\pm 1,\pm 1)/\sqrt{2}$, rather than 2, the degree of the equation.

\end{document}